\documentclass{article}

\usepackage{amsmath} 
\usepackage{amssymb} 
\usepackage[mathcal,mathscr]{eucal} 
\usepackage{eufrak} 
\usepackage{epsfig} 
\usepackage{amsthm} 
\usepackage{graphicx} 
\usepackage[numbers, sort&compress]{natbib} 
 
\newcommand{\tr}{\textrm{tr}} 
\newcommand{\nn}{\nonumber} 
\newcommand{\ket}[1]{|#1\rangle} 
\newcommand{\bra}[1]{\langle#1|} 
 
\newtheorem{theor}{Theor.}[section] 
 
\newtheorem{lemma}{Lemma}[section] 
 
\begin{document} 
 
 
\begin{center}
\Large{\textbf{Some Comments on Three Suggested Postulates for Quantum Theory}}
\end{center}

\begin{center} 
D. Salgado \& J.L. S\'{a}nchez-G\'{o}mez\\ 
Dpto.  F\'{\i}sica Te\'{o}rica, Universidad Aut\'{o}noma de Madrid, Spain \\
\texttt{david.salgado@uam.es} \& \texttt{jl.sanchezgomez@uam.es} \\ 
\end{center}

\medskip


\medskip

\textbf{Journal Reference}: Found. Phys. Lett. \textbf{15}, 209 (2002)

\vspace{0.5cm}

\begin{center}
\textbf{Abstract}

\bigskip

\begin{minipage}{12cm} 
Three basic postulates for Quantum Theory are proposed, namely the Probability, Maximum-Speed and Hilbert-Space postulates. Subsequently we show how these  postulates give rise to well-known and widely used quantum results, as the probability rule and the linearity of quantum evolution. A discussion of the postulates in the light of Bell's theorem is included which points towards yet unsolved conceptual problems in the Foundations of Quantum Mechanics.  
\end{minipage} 
\end{center}

\vspace{1cm} 
 

\section{Introduction} 
 
Special Relativity theory and Quantum Mechanics are usually seen as independent theories. They are even usually viewed as conflicting with each other, especially in the question of the locality/nonlocality nature of Quantum Theory. Despite this possible conflict a ``peaceful coexistence'' between both theories have been explicitly recognized \citep{Shimony}. Here in the same spirit as \citep{Gisinb} we propose three postulates for Quantum Theory among which we include the relativity principle of no faster-than-light speed transmission or alternatively no action-at-distance principle. Thus we strengthen this peaceful coexistence to a \emph{necessary symbiosis} of both theories.

 
\section{Postulates: General View} 
\label{GenView} 
 
First we will very briefly enumerate our proposed postulates and subsequently we will comment on them. We propose the following three postulates as the physical roots of Quantum Theory: 
 
\begin{enumerate} 
\item\label{PP} \textbf{Probability Postulate}.- The physical results predicted by Quantum Theory are probabilistic in nature, i.e. an observer only knows the probability of outcomes in any measurement done upon a quantum system. 
 
\item\label{MSP} \textbf{Maximum-Speed Postulate}.- Physical phenomena, whatever they are, show an upper bound in their transmission speed. 
 
\item\label{HSP} \textbf{Hilbert-Space Postulate}.- It corresponds to every physical system, \textit{whether} simple \emph{or} compound, a Hilbert space, different operators upon which represent the state of the system and the physical quantities to be measured. 
\end{enumerate} 
 
It is obvious that the third one lacks of the same clear physical content as the other two. In our opinion, this responds to the unconlusive question of the definitive interpretation of Quantum Theory and suggests that advancement in this direction should necessarily clarifies the physical meaning behind the Hilbert-space formalism. Notice that the projection postulate has not been included. We firmly believe this postulate not to be a milestone in the physics behind Quantum Theory. 
 
\subsection{Probabilistic Postulate} 
 
The probability nature of Quantum Theory has been one of its most outstanding features since its very creation. Here we will show how this distinguishing feature, far from being just a philosophical question, is fundamental in its mathematical formalism, namely in the evolution of quantum systems. In particular we will show in section \ref{ImmRes} how probability plays a fundamental role to establish the linearity of quantum deterministic evolution. 
 
\bigskip 
 
For probability formalism to be correctly applied we must make sure that Kolmogorov's axioms are satisfied, something which has already been done \citep{BallProb}. This enables us to use well-known probabilistic concepts and results as, in particular, conditional probability and the theorem on compound probabilities \citep{Feller}.  
 
\subsection{Maximum-Speed Postulate} 
 
This postulate is one of the milestones of the theory of Relativity, particularly of the theory of Special Relativity. In this theory  it has fundamental mathematical consequences, namely, the Lorentz transformations between inertial frames, a keystone both in electrodynamics and in any theory incorporating relativity principles.  
 
\bigskip 
 
It should be remarked that this Maximum-Speed postulate also enters in the foundations of Special Relativity, so there's quantum-independent evidence to claim that it is a firmly established physical principle. We very shortly include how this postulate appears in the deduction of Lorentz transformations. If we have two reference frames and investigate what the possible coordinate transformations are under the assumptions of homogeneity and isotropy of space and time and the special relativity principle for inertial frames, we arrive at transformations of the type \citep{Ignatowsky} 
 
\begin{subequations} 
\label{GenLorentz} 
\begin{eqnarray} 
x'&=&\frac{x-Vt}{\sqrt{1-\frac{V^{2}}{\eta}}}\\ 
y'&=&y\\ 
z'&=&z\\ 
t'&=&\frac{t-\frac{V}{\eta}x}{\sqrt{1-\frac{V^{2}}{\eta}}} 
\end{eqnarray} 
\end{subequations} 
 
where $\eta$ is a positive constant with $[Length]^{2}[Time]^{-2}$ dimensions, $V$ is the relative speed of the inertial frames and where we have chosen for concreteness the X axis for the relative direction between the two frames. Once established these coordinate transformations it is straightforward to realize that the existence of an upper bound for the velocity of any physical system, say $c$, determines this constant \citep{Ignatowsky}: 
 
\begin{equation} 
\eta=c^{2} 
\end{equation}  
 
Notice that this approach to Lorentz transformations not only singles out the role played by the Maximum-Speed postulate but also do place Galilean transformations under the same conceptual basis, the only difference being this postulate. In Newtonian space-time it is assumed that there's no upper limit for the velocity, so $\eta=\infty$ and we recover from (\ref{GenLorentz}) the usual Galilean transformations. 
 
\bigskip 
 
There has been previous work in the direction of using this Maximum-Speed principle as a fundamental postulate in Quantum Mechanics, in particular to demonstrate that it forces the quantum deterministic evolution to be necessarily linear \citep{Gisinb,Gisina}. Here we will argue that such a proof is indeed inconsistent, but nevertheless this postulate can shed some light in the locality/nonlocality question posed by Bell's theorem and provide some insight into EPR theorem and EPR elements of reality \citep{EPR}. 
 
\subsection{Hilbert-Space Postulate} 
\label{subHSP} 
This is beyond doubt the more mathematical postulate of the ones we propose here. There's no unanimous consensus about its physical meaning, this being the origin of the different interpretations of the quantum formalism. Instead of assuming any of these interpretations we have included it in its mathematical form in order not to subrestipciously introduce hidden physical assumptions. 
 
\bigskip 
 
This postulate is two-fold. Firstly it is usually assumed that the \emph{state} of quantum systems is represented by vectors in a Hilbert space. We will only claim that there exists a mathematical object related to the Hilbert space which describes the state of the system. On the other hand operators defined upon the Hilbert space are associated with the physical quantities to be measured. These operators are usually called \textit{observables} whereas the measurable physical quantities (naturally assumed as real numbers) are the elements of the spectrum of these observables. Since the spectrum must be real, there must exist a spectral resolution of the identity associated to each of these operators, thus they are selfadjoint. Notice how this formalim naturally includes the possibility of physical quantities having discrete values, a characteristic feature of Quantum Mechanics. 
 
\bigskip 
 
Notice that as it has already been remarked \citep{d'Espagnat} the Hilbert space appears as the fundamental element. Not restricting ourselves to state vectors presents a double advantage: on one hand it enables us to deal with subsystems of compound quantum systems and on the other hand it also allows us to embrace statistical mixtures (ensembles) of quantum systems under the same mathematical formalism.

\section{Postulates: Immediate Results} 
\label{ImmRes} 
 
The choice of these postulates is justified with the following immediate results which are readily obtained from them. 
 
\subsection{Probability Rule: Gleason's Theorem} 
 
In standard textbooks (cf. e.g. \citep{Cohen-Tannoudji}), besides adopting vectors in Hilbert space as representing quantum systems it is also included as a postulate of Quantum Mechanics that the probability of obtaining the result $a_{k}$ in measuring the observable $A$ upon a quantum system in state $\ket{\psi}$ is $Pr(a_{k};\psi)=|\langle a_{k}|\psi\rangle|^{2}$, where $\ket{a_{k}}$ represents the eigenstate associated to the eigenvalue $a_{k}$ of the observable $A$. But as a matter of fact this is not a postulate, since by Gleason' theorem \citep{Gleason} we know that every measure $\mu$ associated to an observable $A$ has the representation $\mu(a_{k})=\tr(\rho P_{k})$ where $\rho$ is a density operator (positive, selfadjoint, unit-trace operator) and $P_{k}$ is a projector associated to $a_{k}$ ($P_{k}=\ket{a_{k}}\bra{a_{k}}$). In particular, if we are interested in defining a probability measure upon the Hilbert space associated to a quantum system we will always be able to find a density operator $\rho$ such that  
 
\begin{equation} 
Pr(a_{k})=\tr(\rho P_{a_{k}}) 
\end{equation} 
 
This theorem, in our opinion, clarifies different fundamental aspects of the widely used quantum formalism. First the density operator $\rho$ is the mathematical object associated to the Hilbert space of a quantum system which was referred to in section \ref{subHSP}. Secondly, notice that the use of the density operator formalism embraces the state vector representation as a particular case. This is easily shown by noting that if $\rho=\ket{\psi}\bra{\psi}$ then  
 
\begin{eqnarray}  
Pr(a_{k})&=&\tr(\ket{\psi}\bra{\psi} P_{a_{k}})=\nn\\ 
&=&\tr(\ket{\psi}\bra{\psi}\ket{a_{k}}\bra{a_{k}})=\nn\\ 
&=&|\langle a_{k}|\psi\rangle|^{2} 
\end{eqnarray}      
 
So that instead of using $\rho$ we may use $\ket{\psi}$ to represent the state of the system. The choice between $\rho$ and $\ket{\psi}$ is particularly irrelevant for pure systems (cf. below), since there's no difference in the physical predictions. Notice that for the time being we are not taking under consideration statistical mixtures of states, so $\rho$ represents a particular state of the system. 
 
\bigskip 
 
Be aware that we are using Gleason's theorem in its original sense (cf. appendix), i.e. to show that you only have to postulate the existence of a measure (probability in this case --Postulate \ref{PP}) upon a Hilbert space (the one associated to each physical system --Postulate \ref{HSP}) to arrive at a density operator. Thus the well-known criticism to the proof of the impossibility of dispersion-free states by resorting to Gleason's theorem \citep{Bell} does not apply here, since any state must be expressed using exclusively the Hilbert-space machinery (Postulate \ref{HSP}).  
 
\bigskip 
 
The advantage of using $\rho$ is double. First it allows within the same mathematical formalism the analysis of ensembles of states. This is due to the linearity of both the trace and the multiplication operations. Suppose that a system is known to be in state $\rho_{k}$ with probability $q_{k}$ $(k=1,\dots,n)$. Obviusly $\sum_{k}q_{k}=1$. Then the probability of finding the value $a_{j}$ of the observable $A$ upon a measurement on this system will be $\tr(\rho_{k}a_{j})$ with probability $q_{k}$, so applying the theorem on compound probabilities we get 
 
\begin{equation} 
Pr(a_{j})=\sum_{k=1}^{n}q_{k}\tr(\rho_{k}a_{j}) 
\end{equation} 
 
from which by linearity we may write 
 
\begin{equation} 
Pr(a_{j})=\tr\left[\left(\sum_{k=1}^{n}q_{k}\rho_{k}\right)a_{j}\right] 
\end{equation} 
 
Since $\sum_{k=1}^{n}q_{k}\rho_{k}$ is selfadjoint, positive and of unit trace, it is a density operator. The difference with the previously considered density operators stems from the fact that $\rho$'s representing statistical mixtures are not idempotent, i.e. 
 
\begin{equation} 
\rho^{2}\neq\rho 
\end{equation} 
 
whereas $\rho$'s of type $ \rho=\ket{\psi}\bra{\psi}$ are. The nonidempotent density operators represents mixture states whereas the idempotent ones are said to represent pure states.  
 
\bigskip 
 
The second advantage appears in the study of compound systems. As we have stated in the Hilbert-space postulate any physical system, \emph{whether simple or compound}, is associated to a Hilbert space. So as it has been stated in section \ref{subHSP}  we must associate a Hilbert space to a, say, double-compound system $\mathcal{H}_{12}$ and its corresponding observables are selfadjoint operators upon $\mathcal{H}_{12}$. But compound means made up of simpler entities, which following the same postulate should be associated to Hilbert spaces $\mathcal{H}_{1}$ and $\mathcal{H}_{2}$, so it is natural to use the tensor product to build up the Hilbert space $\mathcal{H}_{12}$ associated to the compound system:  $\mathcal{H}_{12}=\mathcal{H}_{1}\otimes\mathcal{H}_{2}$, since the tensor product conserves both the linear structure and the scalar product necessary to have a Hilbert space. 
 
\bigskip 
 
Now the advantage of density operators over state vectors is rooted in the fact that whereas in the latter formalism it is impossible to find a state vector describing the state of one of the subsystems in physically interesting situations (product density operators usually appear only as initial conditions and as soon as both subsystems interact we are in the situation considered here), using density operators it is always possible to find another density operator describing the state of a subsystem. This is achieved by resorting to the well-known partial trace operation: 
 
\begin{equation} 
\rho_{1}=\tr_{2}\rho_{12}\qquad\rho_{2}=\tr_{1}\rho_{12} 
\end{equation} 
 
\bigskip 
 
Finally the derivation of the quantum probability rule has also been attempted in the context of particular interpretations of Quantum Mechanics, but these efforts have already been proven to be wrong \citep{SanchGom}, Gleason's theorem being a fundamental result in this respect. The necessity to introduce a probability measure is \emph{inevitable} (cf. \citep{SanchGom}), something which we have assumed in postulate \ref{PP}. 
 
\bigskip 
 
Notice that all these considerations have been made only resorting to postulates \ref{PP} and \ref{HSP}, no further physics has been introduced beyond the one contained in these assumptions. 
 
\subsection{Linear Deterministic Evolution} 
\label{LinEvol} 
 
So far we have not introduced any axiom relating to the dynamics of quantum systems. In this section we will consider the restrictions upon this dynamics imposed by the previous postulates. To study the dynamics of a quantum system we define a family of (super)operators $\{\Phi_{t}\}$ with $t\geq 0$ that carries the state of a system from a state $\rho_{0}$ at an initial time $t_{0}$ to a state described by $\rho_{t}$ at time $t$, i.e. $\Phi_{t}(\rho_{0})=\rho_{t}$. Then it can be proven the following 
 
\begin{theor} 
Under postulates \ref{PP} and \ref{HSP}, the family of superoperators  $\{\Phi_{t}\}_{t\geq 0}$ denoting the dynamics of a quantum system satisfies the relation 
\begin{equation} 
\Phi_{t}(q_{1}\rho_1+q_{2}\rho_{2})=q_{1}\Phi_{t}(\rho_{1})+q_{2}\Phi_{t}(\rho_{2}) 
\end{equation} 
for every convex linear combination $q_{1}\rho_1+q_{2}\rho_{2}$ of idempotent density operators $\rho_{1}$ and $\rho_{2}$. 
\end{theor} 
 
\begin{proof} 
Let's proceed by parts. Let the quantum system described by $\rho$ be a \textbf{closed} system.  Postulate \ref{PP} states that the physical information we obtain from a quantum system is the probability of getting a certain (eigen)value of an observable, say $A$. If the system is in a state  $q_{1}\rho_1+q_{2}\rho_{2}$, then by postulate 3 and Gleason's theorem the probability of measuring the value $a_{j}$ at the initial time is $\tr[(q_{1}\rho_1+q_{2}\rho_{2})P_{a_{j}}]$ and at a time $t$ is  $\tr[\Phi_{t}(q_{1}\rho_1+q_{2}\rho_{2})P_{a_{j}}]$. Now this $\rho$ always \textbf{admits} an ensemble interpretation, i.e. it may always possibly be understood that the system is in state $\rho_{1}$ with probability $q_{1}$ and in state $\rho_{2}$ with probability $q_{2}$. This is proven in the following 
 
\begin{lemma} 
$q_{1}\rho_1+q_{2}\rho_{2}$ always admits an ensemble interpretation for any $\rho_{1}$ and $\rho_{2}$. 
\end{lemma} 
\begin{proof} 
As it has been showed in the preceding section a statistical mixture of states $\rho_{1}$ with probability $q_{1}$ and $\rho_{2}$ with probability $q_{2}$ is represented by the density operator $q_{1}\rho_1+q_{2}\rho_{2}$ whatever $\rho_{1}$ and $\rho_{2}$ are. On the contrary given $q_{1}\rho_1+q_{2}\rho_{2}$  we cannot conclude that it represents a statistical mixture, but only that \emph{within the set of all possible interpretations, the ensemble interpretation must be present}.  This stems out from the fact that you can always construct an ensemble of two arbitrary states $\rho_{1}$ and $\rho_{2}$ with arbitrary probabilities $q_{1}$ and $q_{2}$. Finally it must be remarked that all these deductions only resort to postulates \ref{PP} and \ref{HSP}.  
\end{proof} 
 
Now under the ensemble interpretation the probability of getting the value $a_{j}$ after a time $t$ is given by the theorem on compound probabilities and the linearity of the trace operation $p_{t}(a_{j})=\tr[(q_{1}\Phi_{t}(\rho_1)+q_{2}\Phi_{t}(\rho_{2}))P_{a_{j}}]$ for any $P_{a_{j}}$. We then conclude that $\Phi_{t}(q_{1}\rho_1+q_{2}\rho_{2})=q_{1}\Phi_{t}(\rho_1)+q_{2}\Phi_{t}(\rho_{2})$ under this interpretation.\\ 
Let now the evolution of the quantum system described by $\rho$ be non-linear, i.e. $\Phi_{t}(q_{1}\rho_1+q_{2}\rho_{2})\neq q_{1}\Phi_{t}(\rho_1)+q_{2}\Phi_{t}(\rho_{2})$. Then it is clear that $\rho$ does not admit the ensemble interpretation in clear contradiction to the previous lemma. Thus the evolution for a closed quantum system must be linear. 
 
\bigskip 
 
Let now the quantum system described by $\rho$ be an \textbf{open} system. Then as it is proven in \citep{Gisina, Hughston} its corresponding density operator $\rho\equiv\rho_{S}$ can always be obtained by enlarging its Hilbert space $\mathcal{H}_{S}$ adjoining an auxiliary Hilbert space $\mathcal{H}_{aux}$ in such a way that $\rho_{S}=\tr_{aux}\rho_{S,aux}$, where $\rho_{S,aux}$ is the density operator corresponding to $\mathcal{H}_{S}\otimes\mathcal{H}_{aux}$. This settles the impossibility of distinguishing proper from improper mixtures \citep{d'Espagnat} by local operations. The adjoining is made in such a way that the system plus the auxiliary system can be consider a closed system. Then any family of evolution operators $\Phi_{t}^{S}$ defined over $\mathcal{H}_{S}$ can be obtained by tracing out over the auxiliary Hilbert space $\mathcal{H}_{aux}$: 
 
\begin{equation} 
\Phi_{t}^{S}(\rho_{S})=\tr_{aux}[\Phi_{t}(\rho_{S,aux})] 
\end{equation} 
 
So we can apply the previous result to $\Phi_{t}$, then by the linearity of $\Phi_{t}$ and $\tr_{aux}$ we have \\ 
 

\begin{equation} 
\Phi_{t}^{S}(q_{1}\rho_{S,1}+q_{2}\rho_{S,2})=q_{1}\Phi_{t}^{S}(\rho_{S,1})+q_{2}\Phi_{t}^{S}(\rho_{S,2}) 
\end{equation} 
\end{proof} 
 
A few comments should be made. For the proof to hold it is essential to convince oneself about the existence of at least one closed quantum system. From a conceptual basis this is a delicate question. For instance, the program of decoherence recognizes the notion of open system as fundamental in the analysis of the evolution of any quantum system \citep{Giulini}. But with relation to the foundations of Quantum Theory, two comments must be remarked. First even in decoherence program a closed system must be invoked for the quantum formalism to be applied, i.e. to use Schr\"{o}dinger equation we must have a closed system. So to deny the existence of closed systems invalidates the possibility of using Schr\"{o}dinger equation. Second by definition there's at least one closed system, namely the Universe. This second argument can be refuted by denying the possibility of using Quantum Theory to describe the evolution of the Universe, which amounts to restricting the validity of Quantum Theory, something beyond its present status. 
 
\bigskip 
 
Note that we do not endow the density operator with any particular interpretation. Only the possibility of the ensemble interpretation is necessary, the necessity of it stemming out from the fact that one can \emph{always} construct an ensemble of two arbitrary states with arbitrary probabilities.  
 
\bigskip   
 
Finally in \citep{Gisina} an alternative proof for this same result is given, which seems to violate postulate \ref{MSP}. In particular Gisin claims that ``by measuring [observables] $A$ or $B$ on the system represented by the Hilbert space $\kappa$ [$\mathcal{H}_{aux}$ in our notation], one forces the system represented by $H$ into [one mixture or another]''. But as we will argue in the next section, any local action we make upon the auxiliary system will have a null effect on the system $S$, due to postulate 2. Note also that if postulate \ref{HSP} is changed, this result does not necessarily follow. For instance, in stochastic models (cf. e.g. \citep{Ghirardi}) the evolution might be nonlinear. Here we have shown how the probabilistic character of quantum predictions along with the expression of a quantum state through the Hilbert-space formalism suffices to restrict the evolution of a quantum system, leaving as the only possibility that the latter be linear.

\subsection{Maximum-Speed Postulate: Criticizing EPR Elements of Reality and Reunderstanding Bell's Theorem} 
 
\subsubsection{Marginal and Conditional Probabilities} 
 
\begin{figure}[htbp] 
\begin{center}
\epsfig{file=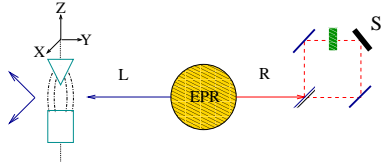, width=5cm}
\caption{\label{EPRfig} Ideal experimental setup for Maximum-Speed postulate discussion.} 
\end{center} 
\end{figure}

To illustrate how this postulate enters into a typical quantum-mechanical situation we will discuss the following ideal experimental setup which is used as an exercise in \citep{BallBook} and which is very close to the arbitrarily fast telephone line built in \citep{Gisinc}. Cf. fig. \ref{EPRfig}. 
 
 
An EPR source sequentially produces pairs of spin $1/2$ particles in singlet state. Then the left ones are directed to a Stern-Gerlach apparatus which can be oriented either along the Z or the X axis at a local observer's will. The other particles are left to travel to a spatially separated region of the former where it will enter a Mach-Zender interferometer with a spin-flipper in its upper arm and in which the first beam-splitter works as a semitransparent mirror if the particle has positive or negative spin component along the Z axis, as a fully transparent glass for particles with negative spin component along X axis and an opaque mirror for particles with positive spin component along X axis: 
 
\begin{eqnarray} 
\label{z+}\ket{z\pm}&\to&\frac{1}{\sqrt{2}}[\ket{\textrm{upper beam}}\pm\ket{\textrm{lower beam}}]\\ 
\label{x-}\ket{x-}&\to&\ket{\textrm{lower beam}}\\ 
\label{x+}\ket{x+}&\to&\ket{\textrm{upper beam}} 
\end{eqnarray}

 A bunch of pairs as numerous as to build an interference pattern is produced in the EPR source and  all the left partners are subsequently measured by a local observer with the Stern-Gerlach magnet oriented along one particular axis. The right partners are then left to enter the interferometer. It is straightforward to realize that if the particles entering the interferometer have positive Z components, then an interference pattern should be expected on the screen S and analogously if they have negative Z component (cf. fig. \ref{EPRfig}). On the contrary if they have components along the X axis, no interference pattern should appear on S. Thus the interferometer may be naively expected to be used to discern whether the bunch of particles R have spin component along the Z or along the X axis. Now due to spin conservation and the fact that each pair has been prepared in a singlet state ($S=0$), if particle R for each pair has spin component along Z then the corresponding particle L will have opposite spin component along Z and the same for the X axis. Apparently (and incorrectly as we will show) we can conclude that if the left observer chooses to measure the, say, Z components of a bunch of particles L, then inevitably an interference pattern will show up on S and the observer of this screen can immediately infer which the choice of his left partner was (to measure along Z or X axis), even if they are spatially separated  and  do not communicate with each other.  
 
\bigskip 
 
We will discuss how postulates \ref{PP} and \ref{MSP} prevents the interference pattern from appearing on the screen, thus also preventing the right observer from inferring along which axis  the left observer chose to measure. We argue that this lack of interference  reflects the fact that the data obtained by the interferometer (provided there is no exchange of information) enables the right observer only to detect marginal probabilities  and not conditional probabilities because the particles entering the interferometer carry no information about the result of the left measurement, this being a consequence of postulate \ref{MSP}.  
 
\bigskip 
 
The argument firstly makes use of postulate \ref{PP} by claiming that the only physical quantity to be measured by the interferometer is the probability that the particle entering it have positive or negative spin components along a particular axis. Now since usual concepts of probability theory are to be applied \citep{BallProb} and  since an observer at the right region is spatially separated from the left one,  this probability can only be chosen out of two possible probability measures, namely the marginal probability and the conditional probability (cf. \citep{Feller}), thus joint probabilities are excluded. The first one expresses the probability of getting $\sigma_{z}^{R}=\pm 1$ irrespectively of the results of the particle L. The second one expresses the probability of getting such physical values conditioned on the results obtained in the left measurement. Analytically, we must make a choice between $Pr(\sigma_{z}^{R})$ and $Pr(\sigma_{z}^{R}|\sigma_{z}^{L})$. Physically this reflects two different circumstances: right particles carrying or not information about the result of the left measurement. Postulate \ref{MSP} enters now into the picture by assuring that, since there's no communication, the data provided by the interferometer can be used only to construct marginal probabilities and not conditional ones. The latter may have been expected since particle L has been measured before particle R enters into the interferometer and one may be inclined to think that it carries information about the result, but it is postulate \ref{MSP} (or some similar version of it) which precludes this possibility from realizing. Note that we do not discuss how to calculate probabilities (whether marginal or conditional) but which ones are to be assigned to the data provided by the interferometer without communication. 
 
\bigskip 
 
Furthermore, postulate \ref{MSP} clearly prevents the interference pattern from appearing on the screen. An interference phenomenon is obtained only if the particles entering the Mach-Zender device have components along the Z axis. But since not having communication it is impossible to know it they have positive or negative spin components (postulate \ref{MSP} once more) the probability of arriving at a point $x$ of the screen will be given by:  
 
\begin{eqnarray} 
Pr(x)&=&Pr(x|\sigma_{Z}^{R}=+1)Pr(\sigma_{Z}^{R}=+1)+\nn\\ 
&+&Pr(x|\sigma_{Z}^{R}=-1)Pr(\sigma_{Z}^{R}=-1) 
\end{eqnarray}  
  
where we have applied the compound probabilities theorem (cf. \citep{Feller}). The conditional probabilities $Pr(x|\sigma_{Z}^{R}=\pm 1)$ are quantum probabilities obtained using eqs. \eqref{z+} and the \emph{a priori} probabilities $Pr(\sigma_{Z}^{R}=\pm 1)=1/2$ since we do not know the result of left measurements. Thus 
 
\begin{eqnarray} 
Pr(x)&=&\frac{1}{2}[|\langle x|\textrm{up.b.}\rangle|^{2}+|\langle x|\textrm{lo.b.}\rangle|^{2}] 
\end{eqnarray}  
  
i.e. no interference is obtained, where

$$\ket{\textrm{up.b.}}\equiv\ket{\textrm{upper beam}}$$  $$\ket{\textrm{lo.b.}}\equiv\ket{\textrm{lower beam}}$$

\bigskip 
 
Note that the Maximum-Speed postulate has been used at three levels. First to recognize that any probability to be measured by an observer makes reference to observables corresponding to that observer. In other words, observer R has no access to joint probabilities, but only to probabilities referring to spin components of particle R. Secondly, this postulate is also used to discern between marginal and conditional probability measures.  Finally to perform calculations of probabilities where quantities referring to (possibly spatially) separated subsystems are involucred. 
 
 
\subsubsection{Internal Consistency and Bell's Theorem} 
 
The appearance of postulate \ref{MSP} as a postulate for Quantum Theory should concern the careful reader who is aware of Bell's theorem. How is it possible that Quantum Theory, not being a local realistic theory (Bell's theorem), incorporates a postulate like that. Thus are the preceding postulates self-consistent? In order to assess this question we will focus on GHZ's formulation of Bell's theorem \citep{GHZ,GHZ2}. We will show how the three previous postulates jointly with the EPR elements of reality are inconsistent. We will settle the theorem step by step so that a further critique could be straightforwardly built upon it. 
 
\bigskip 
 
Let us recall the definition of EPR element of reality \citep{EPR}: ``\emph{If, without in any way disturbing a system, we can predict with certainty (i.e., with probability equal to unity) the value of a physical quantity, then there exists an element of physical reality corresponding to this physical quantity}''. If a quantity $A$ is an element of reality we will denote it as $[A]$. Under the three previous axioms and the latter definition Bell's theorem reads as follows: 
 
\begin{theor}[Bell's theorem] 
The set of postulates \ref{PP}, \ref{MSP} and \ref{HSP} and EPR elements of reality are logically incompatible. 
\end{theor} 
 
\begin{proof} 
Let a  quantum system compound of three spin-1/2 particles be described by the so-called GHZ state  
 
\begin{equation} 
\ket{GHZ}=\frac{1}{\sqrt{2}}\left(\ket{z+z+z+}-\ket{z-z-z-}\right) 
\end{equation} 
 
where henceforth $\ket{n\pm}$ will denote the state with positive/negative spin component along the $n$ axis. 
 
\smallskip 
 
The set of all possible events may be divided into the four disjoint sets of results: 
 
\begin{center} 
\begin{tabular}{c|ccc} 
&&&\\ 
$S_{1}$& $\sigma_{y}^{(1)}\cdot\sigma_{y}^{(2)}=+1$&$\sigma_{y}^{(1)}\cdot\sigma_{y}^{(3)}=+1$&$\sigma_{y}^{(2)}\cdot\sigma_{y}^{(3)}=+1$\\ 
&&&\\\hline\hline 
&&&\\ 
$S_{2}$& $\sigma_{y}^{(1)}\cdot\sigma_{y}^{(2)}=+1$&$\sigma_{y}^{(1)}\cdot\sigma_{y}^{(3)}=-1$&$\sigma_{y}^{(2)}\cdot\sigma_{y}^{(3)}=-1$\\ 
&&&\\\hline\hline 
&&&\\ 
$S_{3}$& $\sigma_{y}^{(1)}\cdot\sigma_{y}^{(2)}=-1$&$\sigma_{y}^{(1)}\cdot\sigma_{y}^{(3)}=+1$&$\sigma_{y}^{(2)}\cdot\sigma_{y}^{(3)}=-1$\\ 
&&&\\\hline\hline 
&&&\\ 
$S_{4}$& $\sigma_{y}^{(1)}\cdot\sigma_{y}^{(2)}=-1$&$\sigma_{y}^{(1)}\cdot\sigma_{y}^{(3)}=-1$&$\sigma_{y}^{(2)}\cdot\sigma_{y}^{(3)}=+1$\\ 
&&& 
\end{tabular} 
\end{center} 
 
Note that these four disjoint sets are equiprobable, thus $P[S_{i}]=+1/4\quad\forall i$. 
 
\smallskip 
 
Let now the three particles be spatially separated from each other so that by postulate \ref{MSP} no physical phenomena can interrelate the three parties, i.e. no possible physical influence from one particle on the others can be established. Then for event $S_{1}$ the following chain of implications can be readily proven 
 
\begin{equation} 
\left.\begin{array}{l} 
P(\sigma_{x}^{(1)}=+1|\sigma_{y}^{(2)}\cdot\sigma_{y}^{(3)}=+1)=1\\ 
P(\sigma_{x}^{(2)}=+1|\sigma_{y}^{(1)}\cdot\sigma_{y}^{(3)}=+1)=1\\ 
P(\sigma_{x}^{(3)}=+1|\sigma_{y}^{(1)}\cdot\sigma_{y}^{(2)}=+1)=1 
\end{array} 
\right\}\overset{(a)}\Rightarrow\left\{\begin{array}{l} 
                                                        \left[\sigma_{x}^{(1)}\right]=+1\\ 
                                                        \left[\sigma_{x}^{(2)}\right]=+1\\ 
                                                        \left[\sigma_{x}^{(3)}\right]=+1 
                                        \end{array}\right\}\overset{(b)}\Rightarrow\left[\sigma_{x}^{(1)}\cdot\sigma_{x}^{(2)}\cdot\sigma_{x}^{(3)}\right]=+1 
\end{equation} 
 
where to calculate probabilities postulates \ref{PP} and \ref{HSP} have implicitly been used (see above), to establish implication (a) the definition of EPR element of reality has been applied and where implication (b) elementarily follows from its premises. Equally for events $S_{2}$, $S_{3}$ and $S_{4}$ similar chains can be posed: 
 
\begin{equation} 
\left.\begin{array}{l} 
P(\sigma_{x}^{(1)}=+1|\sigma_{y}^{(2)}\cdot\sigma_{y}^{(3)}=+1)=1\\ 
P(\sigma_{x}^{(2)}=-1|\sigma_{y}^{(1)}\cdot\sigma_{y}^{(3)}=-1)=1\\ 
P(\sigma_{x}^{(3)}=-1|\sigma_{y}^{(1)}\cdot\sigma_{y}^{(2)}=-1)=1 
\end{array} 
\right\}\overset{(a)}\Rightarrow\left\{\begin{array}{l} 
                                                        \left[\sigma_{x}^{(1)}\right]=+1\\ 
                                                        \left[\sigma_{x}^{(2)}\right]=-1\\ 
                                                        \left[\sigma_{x}^{(3)}\right]=-1 
                                        \end{array}\right\}\overset{(b)}\Rightarrow\left[\sigma_{x}^{(1)}\cdot\sigma_{x}^{(2)}\cdot\sigma_{x}^{(3)}\right]=+1 
\end{equation} 
 
\begin{equation} 
\left.\begin{array}{l} 
P(\sigma_{x}^{(1)}=-1|\sigma_{y}^{(2)}\cdot\sigma_{y}^{(3)}=-1)=1\\ 
P(\sigma_{x}^{(2)}=-1|\sigma_{y}^{(1)}\cdot\sigma_{y}^{(3)}=-1)=1\\ 
P(\sigma_{x}^{(3)}=+1|\sigma_{y}^{(1)}\cdot\sigma_{y}^{(2)}=+1)=1 
\end{array} 
\right\}\overset{(a)}\Rightarrow\left\{\begin{array}{l} 
                                                        \left[\sigma_{x}^{(1)}\right]=-1\\ 
                                                        \left[\sigma_{x}^{(2)}\right]=-1\\ 
                                                        \left[\sigma_{x}^{(3)}\right]=+1 
                                        \end{array}\right\}\overset{(b)}\Rightarrow\left[\sigma_{x}^{(1)}\cdot\sigma_{x}^{(2)}\cdot\sigma_{x}^{(3)}\right]=+1 
\end{equation} 
 
\begin{equation} 
\left.\begin{array}{l} 
P(\sigma_{x}^{(1)}=-1|\sigma_{y}^{(2)}\cdot\sigma_{y}^{(3)}=-1)=1\\ 
P(\sigma_{x}^{(2)}=+1|\sigma_{y}^{(1)}\cdot\sigma_{y}^{(3)}=+1)=1\\ 
P(\sigma_{x}^{(3)}=-1|\sigma_{y}^{(1)}\cdot\sigma_{y}^{(2)}=-1)=1 
\end{array} 
\right\}\overset{(a)}\Rightarrow\left\{\begin{array}{l} 
                                                        \left[\sigma_{x}^{(1)}\right]=-1\\ 
                                                        \left[\sigma_{x}^{(2)}\right]=+1\\ 
                                                        \left[\sigma_{x}^{(3)}\right]=-1 
                                        \end{array}\right\}\overset{(b)}\Rightarrow\left[\sigma_{x}^{(1)}\cdot\sigma_{x}^{(2)}\cdot\sigma_{x}^{(3)}\right]=+1 
\end{equation} 
 
Now since these four cases (each chain corresponds to one $S_{i}$) exhaust all possibilities, we arrive at the conclusion that with probability $1$ the quantity $\sigma_{x}^{(1)}\cdot\sigma_{x}^{(2)}\cdot\sigma_{x}^{(3)}$ must have an element of reality, and its value must be  
 
\begin{equation} 
\left[\sigma_{x}^{(1)}\cdot\sigma_{x}^{(2)}\cdot\sigma_{x}^{(3)}\right]=+1 
\end{equation} 
 
But applying quantum formalism (postulates \ref{PP} and \ref{HSP}) we may also calculate $P[\sigma_{x}^{(1)}\cdot\sigma_{x}^{(2)}\cdot\sigma_{x}^{(3)}=+1]=0$, since $P[\sigma_{x}^{(1)}\cdot\sigma_{x}^{(2)}\cdot\sigma_{x}^{(3)}=-1]=1$. Thus the hypotheses are self-contradictory.  
\end{proof} 
 
It should be remarked that this same scheme can be applied to Hardy's \citep{Hardy} and Cabello's \citep{Cabello} versions of Bell's theorem. 

\subsubsection{Analysis of Bell's theorem} 
 
How is it then possible to overcome this contradiction? Are we necessarily  to abandon one of the hypotheses (or even more than one)? The commom view  is to claim that the locality principle (postulate \ref{MSP}) does not hold any more in the quantum realm, though some refinements of the concept of locality may also be found \citep{BallentJarr}. From the previous proof of Bell's theorem it should be clear that the crucial point and more controversial step is the implication (a), implication (b) following elementarily from the conclusions of (a). As a matter of fact, if locality drops out and a possibility of mutual influence (though strange as it may be) is established, then (a) does not follow and the inconsistency wipes out. This is the typical situation in which a reduction postulate is taken into account. The influence is settled in such a way that no information about the result can be recovered by local measurements upon the spatially separated influenced partner. We address here instead the possibility of precluding implication (a) by an epistemological distinction in the definition of EPR elements of reality. 
 
\bigskip  
 
The epistemological distinction we propose to introduce in the definition of EPR elements of reality concerns the concept of probability to be used to establish the certainty of the prediction of a physical quantity. Following a minimal quantum formalism (postulates \ref{PP} and \ref{HSP}) it is clear that in general a quantum system does not possess any of its possible results prior to a measurement upon it (cf. \citep{Mermin, Mittel, Redhead}), so that the final value registered by the measuring device is the product of the interaction between the system and the apparatus. In addition since all known interactions are local \citep{Weinberg} this apparatus-system interaction must also be local, thus we only have access to local information. This, as it have been argued before, is revealed by the fact that probabilities directly accesible to the experiment are marginal probabilities never conditional ones and this is theoretically justified by postulate \ref{MSP}. We want to stress the fact that this locality does not preclude quantum correlations like e.g. for the GHZ state. We are thus left with a highly paradoxical situation: on one hand all interactions in Nature are local even in the quantum realm and on the other hand correlations are present even when the correlated values for spatially separated systems are established through local interactions with measuring apparatus. In these sense these so-called quantum correlations are nonlocal. We believe that this locality/nonlocality dualism can help to better understand the inconsistency previously depicted. 
 
\bigskip 
 
The main idea is to realize that marginal probabilities are more fundamental from a rigorous epistemological point of view than conditional probabilities. Again we want to stress the fact that this does not invalidate the latter, it only partially reduces its epistemological value. Note that this distinction parallels the previous locality/nonlocality dualism. Fundamental locality is revealed through marginal probabilities, i.e. probabilities directly accessible to the experiment whereas the nonlocal character of quantum correlations is established using conditional probabilities. If we want to formulate a notion of how physical processes take place in Nature, would it not be more natural to limit this notion to information directly accesible to the experiment? In some approximate  sense this is analogous to the situation relating Maxwell's demon and classical entropy (cf. e.g. \citep{Sklar}): if an entity such as Maxwell's demon existed then in principle it would be possible to decrease entropy, though the non-decreasing of entropy is a fundamental law of Nature (for many the most fundamental one). Note that the existence of this entity would only require to be able to distinguish between particles with high and low velocities. As a matter of fact a well-trained person (if he/she were capable of seeing particles) could perform this task very easily and however a closer analysis (see \citep{Sklar} for references) concludes that this does not imply that entropy is actually decreased in fundamental processes in Nature, i.e. Nature always works not decreasing entropy.  
 
\bigskip 
 
In typical situations as the ones depicted before it is clearly possible to send the information obtained in one party over to the other(s) and the latter can process this information to identify which events where conditioned by which previous results. Thus in principle judging by the correlations it is possible to establish an action-at-a-distance influence between both parties. But as in the Maxwell's demon paradox a closer analysis would reveal that interactions are always local, thus neglecting the existence of the alleged action at a distance. Fundamental processes, i.e. those accesible to experiments tell us that Nature always works locally. Notice once more that we are not neglecting the existence of correlations, we only criticize the way of understanding how physical processes take place in Nature with the definition of EPR elements of reality. 
 
\bigskip 
 
Analytically the distinction between the locality by which physical process operate and the nonlocality of quantum correlations is proposed to be rooted upon the difference between marginal and conditional probabilities. Thus if conditional probabilities are used in the definition of EPR elements of reality (conditional probability equals 1), we are reviving a Maxwell's-demon-like creature   and these elements of reality go beyond how fundamental processes take place in Nature, but if only marginal probabilities are used (those accesible to local experiments), then we may find a way out of Bell's theorem conclusion, since implication (a) is not justified.

\bigskip 
 
Note that neither this is a complete negation of Bell's theorem nor a negation of quantum correlations. It is only an epistemological remark trying to single out what shocking and puzzling features of Quantum Mechanics remain to be well understood. How nonlocal correlations are established through spatially separated acts of measurements prior to which no value can be assigned to each party remains a mystery (at least for us). The sexagenary proposed reduction postulate, an undetectable process with a completely unknown dynamics which clearly violates Special Relativity (if it is to be considered a physical process taking place in space-time), does not seem to us the definitive solution and we believe that further efforts both in the theoretical and experimental sides should be made to understand this paradox. 
 
\section{Conclusions} 
 
We have proposed a set of basic postulates for Quantum Theory. Two of these involve clear-cut physical statements upon the nature of the quantities to be predicted by the theory (Probability postulate) and the possible connection among these different quantities (Maximum-Speed postulate). The last one (Hilbert-Space postulate) shows a strong mathematical character, suggesting in our opinion the lack of a definitive physical interpretation. Any attempt towards new interpretations of Quantum Theory should, we believe, clarify the content of this principle in more physical terms. 
 
\bigskip 
 
This choice of postulates has relevant consequences both for the formalism and for its understanding. First Gleason's theorem emerges from these assumed principles in a natural fashion, thus providing the fundamental tool to express the state of a quantum system, namely the density operator. Second the linearity of quantum evolution follows from the probability and Hilbert-space postulates without resorting to any other physical assumption. Hence probability and Hilbert space formalisms appear to be more restrictive than what  can be thought at first glance. Finally, similarly to previous proposal (cf. \citep{Gisinb}), a locality criterion (Maximum-Speed postulate; already present in Special Relativity) is proposed to enter into this basic set of postulates.

\bigskip  
  
After proving Bell's theorem from these postulates and the well-known definition of EPR elements of reality, we have discussed how an epistemological distinction between marginal and conditional probabilities can be introduced in the identification of the previous elements of reality, thus relaxing Bell's theorem conclusions.  
 
\bigskip 
 
However the establishment of nonlocal quantum correlations remains a mystery under this set of postulates and we conclude that further efforts should be made to understand uncontroversially the process by which these correlations set up.

\section*{Appendix A: Gleason's theorem} 
 
For completeness we include the original statement of Gleason's theorem \citep{Gleason}: 
 
\begin{description} 
\item{\textbf{Gleason's theorem}:} Let $\mu$ be a measure on the closed subspaces of a separable (real or complex) Hilbert space $\mathcal{H}$ of dimension at least three. There exists a positive semi-definite self-adjoint operator $T$ of the trace class such that for all closed subspaces $A$ of $\mathcal{H}$ $$\mu(A)=\textrm{trace}(TP_{A})$$where $P_{A}$ is the orthogonal projection of $\mathcal{H}$ onto $A$. 
\end{description} 
 
Note that the theorem does not make statements about hidden-variable theories or interpretations (cf. \citep{Jammer} for details about the role of Gleason's theorem in the hidden-variables subject). It is a purely mathematical result. 
\\ 
 
 
\section*{Acknowledgments} 
We acknowledge the support of Spanish Ministry of Science and Technology under project No. BFM2000-0013. One of us (D.S.) must also acknowledge the support of Madrid Education Council under grant BOCAM 20-08-1999. We are most grateful to Pr. G. Garc\'{\i}a-Alcaine for his careful reading of the manuscript as well as to an anonymous referee for his useful criticisms.


\end{document}